# Exploiting real-time FPGA based adaptive systems technology for real-time Sensor Fusion in next generation automotive safety systems

Steve Chappell a , Alistair Macarthur a , Dan Preston b , Dave Olmstead b
Bob Flint c & Chris Sullivan a
a Celoxica Ltd, 66 Milton Park, Abingdon, Oxfordshire OX14 4RX, U.K.
b Medius Inc. 911 Western Ave, Suite 530, Seattle, WA 98104, U.S.A.
c BAE SYSTEMS Ventures, Warwick House, Aerospace Centre, Farnborough GU14 6YU, U.K.

## Abstract

*We present a system for the boresighting of sensors using inertial measurement devices as the basis for developing a range of dynamic real-time sensor fusion applications. The proof of concept utilizes a COTS FPGA platform for sensor fusion and real-time correction of a misaligned video sensor. We exploit a custom-designed 32-bit soft processor core and C-based design & synthesis for rapid, platform-neutral development. Kalman filter and sensor fusion techniques established in advanced aviation systems are applied to automotive vehicles with results exceeding typical industry requirements for sensor alignment. Results of the static and the dynamic tests demonstrate that using inexpensive accelerometers mounted on (or during assembly of) a sensor and an Inertial Measurement Unit (IMU) fixed to a vehicle can be used to compute the misalignment of the sensor to the IMU and thus vehicle. In some cases the model predications and test results exceeded the requirements by an order of magnitude with a 3-sigma or 99% confidence.*

## 1. Automotive Sensor Fusion

Next generation automotive systems and Advanced drive assistance systems (ADAS) such as Lane Departure Warning, Collision Avoidance, Blind Spot Detection or Adaptive Cruise Control require the overlay or "*fusion*" of data from sensors such as video, radar, laser, global positioning systems and inertial measurement devices. Techniques for sensor fusion, namely the incorporation of disparate and complementary sensor data in order to enhance accuracy, are well established in advanced aviation systems[1]. Indeed, as automotive systems incorporate ever more advanced sensor electronics, manufacturers will look to new approaches to system design incorporating reconfigurable devices for rapid development and deployment. As with the application example discussed here, sensor fusion in defense and aerospace systems may provide some solutions.

## 2. Boresighting

It is critical for these next generation applications that the sensors used are accurately aligned and continuously realigned to each other and the vehicle. Without alignment the information from one sensor cannot be effectively combined with that from a different sensor. The current state of the art is to use costly mechanical and optical methods, such as autocollimators and laser boresight tools during vehicle production. Moreover, these alignments must be repeated if a sensor is disturbed (e.g. through typical 'car park' bumps) or subsequently repaired/ replaced after production in order to ensure that safety critical systems are reporting accurately. It is therefore highly desirable to align or "boresight" sensors dynamically through digital computation rather than mechanical means. We present the development of a system for this purpose below.

## 3. Application overview

In order to boresight a sensor we use a combination of two other sensor systems:
- A 6 degree-of-freedom (6-DOF) inertial measurement unit (IMU), such as that from BAE Systems[ii], consisting of 3 gyroscopes and accelerometers, fixed to the vehicle and;
- A two-axis accelerometer (ACC) fixed to the sensor to be boresighted.

Fixed to the vehicle the IMU defines the moving platform reference frame. It can sense angular rate changes and linear accelerations relative to the principal axes of the vehicle $(x,y,z)$. The ACC is fixed to the sensor to be aligned and defines the sensor axes $(x',y',z')$. As the vehicle accelerates, the common acceleration vector will be sensed by both the IMU and the ACC (Figure 1).

Figure 1: Sensor and vehicle reference frames

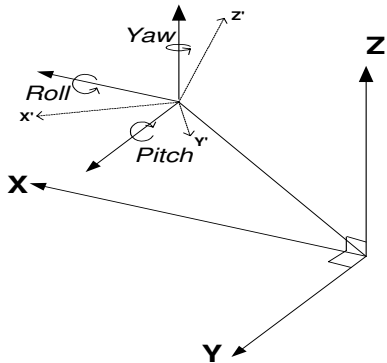

Any differences in acceleration components along the sensor axes are a result of the misalignment between the two and individual instrument errors. We exploit these differences to computationally boresight the sensor.

We calculate sensor misalignment as roll, pitch, and yaw values using a "Sensor Fusion Algorithm"[iii] incorporating a Kalman Filter[iv]. The Kalman Filter also generates a statistical confidence level in the misalignment values. For the purpose of visualization we chose to boresight a video camera as the misaligned sensor, however the method extends to the alignment of any meaningful directional sensor type (e.g. radar, lidar). The misalignment angles are input to an "Affine Transform"[v] to calculate and display a realigned version of the video input in real-time.

Figure 2: System Architecture.

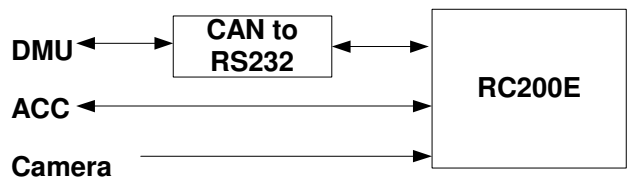

## 4. Sensors

The system incorporates a 6-DOF IMU (DMU) from BAE SYSTEMS[vi] and dual-axis ACC from Analog Devices (ADXL202EB-232A)[vii]. These sensors use Micro-Electrical-Mechanical Systems (MEMS) technology. The DMU incorporates vibrating gyroscopes[viii]. The principle of operation is based on the Coriolis effect whereby, on rotation, secondary vibrations are induced in a vibrating element orthogonal to the original direction of vibration. The rate of angular turn can be detected by measuring these vibrations. The vibrating element in the IMU used consists of a ring resonator micro-machined from silicon. This creates a robust gyroscope able to withstand extreme shock and vibration.

The accelerometers in both IMU and ACC determine acceleration through changes in the capacitance between independent fixed plates and central plates attached to a moving mass

## 5. Sensor Fusion Algorithm

After data reconstruction and subsequent data fusion, the data is passed through a Kalman Filter that tracks the sampled data and provides a confidence level of the tracking quality. The resultant values from this combined sensor fusion algorithm are roll, pitch and yaw of the boresighted sensor with respect to the IMU axes, with associated covariance values, that give an indication of the error in predicted output.

## 6. Affine Transformations

In order to boresight and stabilize the image from the video sensor we take the roll, pitch and yaw values directly from the sensor fusion algorithm. These movement values are then used in co-ordinate transforms for correcting the video picture. These transforms preserve parallel lines and are known as Affine transformations:

$$r' = Ar + B,$$

where **A** is the coordinate rotation matrix for angle $\theta$ about the z axis

$$A = \begin{pmatrix} cos\theta & -sin\theta \\ sin\theta & cos\theta \end{pmatrix}$$

and **B** is the linear transformation vector for corrections $b_x$ and $b_y$ in $x$ and $y$ respectively

$$B = \begin{pmatrix} b_x & 0 \\ 0 & b_y \end{pmatrix}$$

## 7. System Architecture

We use a COTS (Commercial off the Shelf) platform to demonstrate rapid integration and development of the proof of concept prototype. The system is illustrated in Figure 2 and consists of the IMU and ACC sensors plus video camera to be aligned. The IMU interfaces to CAN. The ACC interfaces to Serial. By using a CAN to Serial converter we limit any customisation of the COTS hardware to incorporating a second serial interface onto the chosen platform.

The Celoxica RC200E[ix] was used as the base platform for the prototype. It incorporates a Virtex2 FPGA (XC2V1000 )[x], two banks of 2 Mbyte ZBT RAM, Video I/O, serial interfaces and a TFT display.

Figure 3: FPGA system

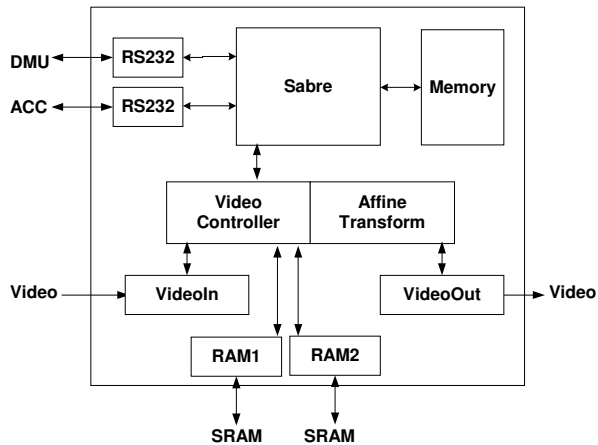

## 8. FPGA System

We exploit the flexibility of FPGAs to rapidly develop and integrate the various system components using the Celoxica DK Design Suite[xi] for C based design and synthesis. The FPGA system contains the sensor interface drivers, the sensor fusion algorithm, video manipulation and user display code. The main architectural components are shown as a schematic in Figure 3 and as code in Figure 4. A platform support library[xii] (PSL) provides ready-coded components for implementation of controllers and data transfer to the board peripherals such as RS232/ video.

For implementation of the main Kalman Filter computation and to control program flow including parts of the Graphical User Interface (GUI) we use the "Sabre" soft-core processor. All other components are implemented directly in the FPGA. In particular the real-time video transformation has intensive processing requirements beyond the capabilities of typical embedded micro and DSP devices.

Figure 4: FPGA system code

```
void main (void)
{
    …

    // Run everything

    par{ // Run Hardware Components
       SabreRun(&MyBus); // 32-bit Processor
       RAMRun(RAM1);     // RAM Framebuffer
       RAMRun(RAM2);     // RAM Framebuffer
       VideoInRun (VideoIn); // Video Input
Stream
       VideoOutRun(VideoOut);// Video Output
Stream

        seq{
           par{ // Enables on Startup
              RAMEnable(RAM1);
              RAMEnable(RAM2);
              VideoInEnable (VideoIn);
              VideoOutEnable(VideoOut);
           }
           // main control loop
           seq{
              WaitForSabre(); // Wait for Kalman Result
              par{
                 // Capture Video
                 VideoInProcess VideoIn);
                 //Affine Transform and Output Video
VideoOutProcess(VideoOut);
              }
           }
        }
    }
}
```

## 9. Video Transformation

Video processing consists of two routines described using Handel-C[xiii], an ANSI-C superset optimized for hardware design in Programmable Logic. VideoInProcess() makes use of the RC200 library routines to take data from the relevant video input device and write successive frames of data to RAM. VideoOutProcess() computes the Affine transformation of coordinates on the RAM framebuffer, copying the relevant pixels to output to the video display. The video processing makes use of both RC200 RAMS in a double-buffering scheme. The routine for the rotation transform is given in Figure 5. This is a five-stage pipeline which, once loaded, computes the rotated output location (OutX,OutY), of each input pixel (InX,InY) on each clock cycle. The transforms operate on 16-bit precision fixed point values with sine and cosine angles stored in a 1024-element lookup table.

```
OutX = InX.cos(theta) − InY.sin(theta)
OutY = InY.cos(theta) + InX.sin(theta)
```

Figure 5: Affine Transformation C code

```
static macro proc
RotateCoordinates(theta,InX,InY,OutX,OutY)
{
    …
    par{
        // Pipeline step 1
        GenerateSine(theta,Sin);
        GenerateCos(theta,Cos);

        //Pipeline step 2
        mapX = InX − CentreOfRotation[0];
        mapY = InY − CentreOfRotation[1];
        temp[0] = Int2fixed(mapX);
        temp[1] = Int2fixed(mapY);

        // Pipeline step 3
```

```
        FixedMult(temp[1], -Sin, temp[2]);
        FixedMult(temp[0], Cos, temp[3]);

        FixedMult(temp[0], Sin, temp[4]);
        FixedMult(temp[1], Cos, temp[5]);

          //Pipeline step 4
        mapXback = fixed2Int(temp[2]+temp[3]);
        mapYback = fixed2Int(temp[4]+temp[5]);

          //Pipeline step 5
        OutX = mapXback + CentreOfRotation[0];
        OutY = mapYback + CentreOfRotation[1];
        }
}
```

## 10. Sabre Processor

Sabre is a 32-bit RISC, designed in Handel-C, and programmed into the FPGA as a soft-core. It has a Harvard architecture[xiv], with expandable data and program memories, limited only by the availability of FPGA embedded block ram. On the VirtexII 1000, there are 80 BlockRams, giving us up to 8kbyte program memory, for instructions and stack, and 64kbyte of data memory for constants. The Sabre is connected via 32-bit buses to each of these memories. Peripherals are simply connected via another 32-bit bus into the processor memory space (see Figures 6 and 7) where the Sabre acts as the bus master. The peripherals are designed to be as "smart" as possible, reducing the workload for the processor and making best use of the parallel processing capabilities of the FPGA. For example, the SabreBusControlRun peripheral simply consists of a set of twelve memory-mapped registers including roll, pitch and yaw values and status flags that are used directly by the FPGA video transformation block. Similarly the SabreBusExpansionPort peripherals connect to the IMU and ACC devices via serial communications blocks.

In this application the serial communications blocks, Sabre processor core, peripherals and board interface support were compiled to a device optimized EDIF netlist using the DK Design Suite. Xilinx place and route tools generated the FPGA configuration file. The Sabre program code was written in C and compiled to the Sabre Instruction Set Architecture. Since the Sabre machine code resides entirely within BlockRam memory of the FPGA, it is a simple process to merge the BlockRam initialisation into the FPGA configuration file. This technique eliminated the need for full hardware recompilation following changes to the Sabre software during development. We could thus rapidly prototype functionality in C *software* and later partition and re-partition functionality into C for *hardware* to improve performance and reduce processor load.

Note that, as a result of the dynamic range of the Kalman filter, it was necessary to use floating-point values for all intermediate stages. The version of Sabre used here has no floating-point co-processor. We therefore emulated IEEE floating point operations using the "Softfloat" library[xv].

Softfloat is widely used amongst RISC processors with no native floating point support, such as low power ARMs and PowerPCs.

Figure 6: Sabre Processor System Architecture

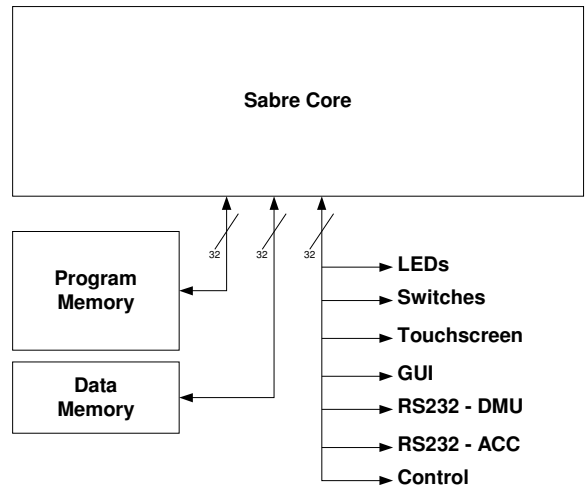

Figure 7: Sabre Processor System C code

```
void SabreRun (BusPtr)
{
   …

   par{

   /* Core components */

       SabreRun           (BusPtr, DATA_MEMORY,
PROGRAM_MEMORY);
       SabreBusRun        (BusPtr);
       SabreBusMemoryRun  (BusPtr,
BUS_BASE_ADDRESS);

   /* User defined Peripherals */

        //LEDs
       SabreBusLEDsRun        (BusPtr,
LEDS_BASE_ADDRESS);
       //Switches
       SabreBusSwitchesRun    (BusPtr,
SWITCHES_BASE_ADDRESS);
       // TouchScreen
       SabreBusTouchScreenRun (BusPtr,
TSCREEN_BASE_ADDRESS);
       // Graphical Output to Screen
```

```
        SabreGuiRun         (BusPtr,
LINE_BASE_ADDRESS, …);
    // AMU Interface
        SabreRS232DMURun       (BusPtr,
SERIAL1_BASE_ADDRESS);
    // DMU Interface
        SabreRS232ACCRun       (BusPtr,
SERIAL2_BASE_ADDRESS);
    // Registers for Affine Transform
        SabreControlRun        (BusPtr,
ANGLES_BASE_ADDRESS);

    }
```

## 11. Testing and Results

Sets of static and dynamic tests were performed to calculate the response and accuracy of the system. In these tests, the system was calibrated first and then misalignments of a few degrees were introduced in roll, pitch and yaw to the boresighted sensor. The correction system was then started and data was collected for 300 seconds. The residuals, the difference between the true accelerometer measurement and the expected acceleration measurement, were used to help tune the Kalman Filter by selecting a good measurement noise value. For the static tests the value could be set very low, about .003 to .01 m/s, since the only noise was the noise of the instruments. This value had to be increased to .015 or higher when the vehicle was moving because of the addition of the vehicle vibration. Figure 8 shows the X-axes residuals and it's 3-sigma value plotted together for a static run and a moving run. The static run shows the residuals well within the 3-sigma values while the moving tests show that the residuals do exceed the 3-sigma values. Since the residuals should only exceed the 3-sigma value about once every 100 samples, the Filter noise was increased.

### 11. 1. Static Testing

The instruments were calibrated using a level test platform. The absolute misalignments were measured directly using a laser attached to the boresighted sensor. Note that static roll and yaw tests are more difficult to perform than the pitch tests since the platform must be oriented and use gravity to generate components of acceleration in the ACC and DMU accelerometers.

Table 1 demonstrates that the resulting alignment estimates were very accurate in all three axes.

Table 1: Results from Static (Top) & Dynamic (Bottom) Tests

| Maneuver Performed | True Angle | Avg Estimated Angle | Filter Confidence |
|---|---|---|---|
| Pitch up | 1 degree | .979 degrees | .011 degrees |
| Pitch down | 1 degree | -1.002 degrees | .011 degrees |
| Roll left | 2 degrees | -2.082 degrees | .011 degrees |
| Roll right | 2 degrees | 1.986 degrees | .011 degrees |
| Yaw left | 1 degree | -1.005 degrees | .012 degrees |
| Yaw right | 1 degree | 1.073 degrees | .012 degrees |

| Test No. | Roll Est. Degrees | Pitch Est. Degrees | Yaw Est. Degrees | Roll SD Degrees | Pitch SD Degrees | Yaw SD Degrees |
|---|---|---|---|---|---|---|
| 1 | -2.152 | -1.598 | -2.389 | .009 | .009 | .079 |
| 2 | -2.199 | -1.572 | -2.224 | .011 | .011 | .087 |

Figure 8: X Axis residuals from Static (Top) and Dynamic (Bottom) Tests.

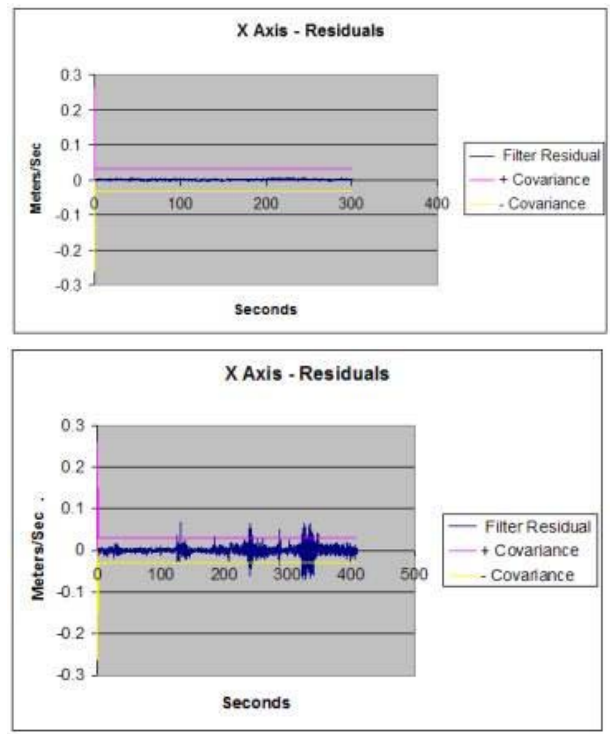

### 11.2. Dynamic Testing

The dynamic tests consisted of placing the test equipment in a standard private passenger vehicle, calibrating the ACC and IMU, misaligning the ACC-Camera system and then running during car motion. Sample results are given in Figure 9.

Note that it is difficult to run precisely the same test profile using a moving vehicle, however the results for two driving tests are shown in Table 1. It can be seen that there is very close agreement between the tests with a high confidence level result.

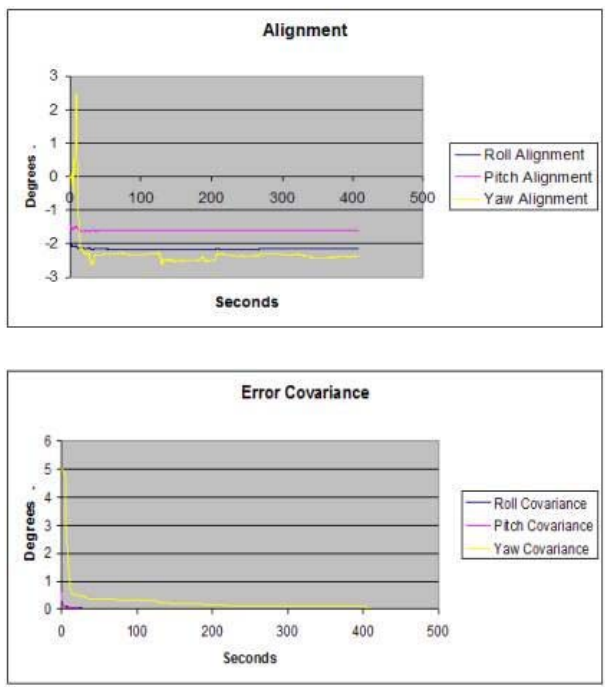

Figure 9: Sample results from dynamic test

## 12 Conclusion

Results of the static and the dynamic tests demonstrate that using inexpensive accelerometers mounted on (or during assembly of) a sensor and an Inertial Measurement Unit (IMU) fixed to the vehicle can be used to compute the misalignment of the sensor to the IMU and thus vehicle. The overall accuracy is dependent on the accuracy of the inertial instruments, mounting accuracy of the instruments, noise present at the sensors and time allowed for the filter to compute the misalignment angles. It is clear that a software based sensor fusion approach to sensor alignment for vehicle applications meets typical automotive industry requirements. In some cases the model predications and test results exceeded the requirements by an order of magnitude with a 3-sigma or 99% confidence. The use of a COTS FPGA board and C based design and synthesis greatly simplified and accelerated the task of creating and verifying a real-time proof of concept system. By abstracting the core design from the details of the platform, the system is highly portable across different platforms and lends itself well to product migration. We note that optimization of the performance (clock-speed, program size etc) was not a design goal in this exercise and there are many obvious enhancements. For example, a full fixed-point analysis and conversion of the Sensor Fusion Algorithm from float to fixed-point calculations is possible. The boresighted sensor system presented here provides the basis for developing a range of dynamic real-time sensor fusion systems. Future implementations will demonstrate self-aligning and self-referencing methods for dynamic alignment of multiple sensors; the fusion of data from the vehicle into the system for additional improvements; and alignment for other sensor features such as headlights. The fusion engine presented here provides self-boresighting functionality for individual sensors, but it can readily be extended to fuse data from multiple sensors together (eg. lidar and video) to provide low-cost situational awareness systems for automotive use. It is proposed these implementations and tests be seriously considered by the automotive industry, as we believe these will aide in the reduction of future system costs, shorten development time to market, facilitate and consolidate the improvement of the next generation of features for the automotive industry.